\documentclass[a4paper,11pt]{article}
\pdfoutput=1 

\usepackage{jcappub} 

\usepackage[T1]{fontenc} 

\usepackage{amssymb}
\usepackage{amsmath}
\usepackage{subcaption}
\usepackage{orcidlink}

\usepackage{hyperref}
\usepackage{amsthm}
\usepackage{comment}

\usepackage[symbol]{footmisc}

\makeatletter
\newcommand{\corref}[1]{\footnote{\textbf{Corresponding author}}}
\makeatother

\makeatletter
\newcommand{\cortext}[2][]{%
  \def\@tempa{#1}%
  \ifx\@tempa\@empty
    \footnotetext{#2}%
  \else
    \footnotetext{\textbf{#1:} #2}%
  \fi
}

\makeatother

\newcounter{daggerfootnote}

\usepackage{lineno}

\begin{document}





\title{Blazars as a Potential Origin of the KM3-230213A Event}


%


\author[b,a]{O.~Adriani\,\orcidlink{0000-0002-3592-0654}}
\author[c,be]{A.~Albert}
\author[d]{A.\,R.~Alhebsi\,\orcidlink{0009-0002-7320-7638}}
\author[d]{S.~Alshalloudi}
\author[e]{S. Alves Garre\,\orcidlink{0000-0003-1893-0858}}
\author[g,f]{A. Ambrosone\footnote{corresponding author}}
\emailAdd{km3net-pc@km3net.de}
\emailAdd{antonio.ambrosone@unina.it}
\author[h]{F.~Ameli}
\author[i]{M.~Andre}
\author[j]{L.~Aphecetche\,\orcidlink{0000-0001-7662-3878}}
\author[k]{M. Ardid\,\orcidlink{0000-0002-3199-594X}}
\author[k]{S. Ardid\,\orcidlink{0000-0003-4821-6655}}
\author[l]{J.~Aublin}
\author[n,m]{F.~Badaracco\,\orcidlink{0000-0001-8553-7904}}
\author[o]{L.~Bailly-Salins}
\author[l]{B.~Baret}
\author[e]{A. Bariego-Quintana\,\orcidlink{0000-0001-5187-7505}}
\author[p]{M.~Barnard\,\orcidlink{0000-0003-1720-7959}}
\author[l]{Y.~Becherini}
\author[f]{M.~Bendahman\footnote{corresponding author}}
\emailAdd{meriem.bendahman@na.infn.it}
\author[r,q]{F.~Benfenati~Gualandi}
\author[s,f]{M.~Benhassi}
\author[t]{D.\,M.~Benoit\,\orcidlink{0000-0002-7773-6863}}
\author[v,u]{Z. Be\v{n}u\v{s}ov\'a\,\orcidlink{0000-0002-2677-7657}}
\author[w]{E.~Berbee}
\author[w]{C.~van~Bergen}
\author[b]{E.~Berti}
\author[x]{V.~Bertin\,\orcidlink{0000-0001-6688-4580}}
\author[b]{P.~Betti\,\orcidlink{0000-0002-7097-165X}}
\author[y]{S.~Biagi\,\orcidlink{0000-0001-8598-0017}}
% ...
\affiliation[a]{Universit{\`a} di Firenze, Dipartimento di Fisica e Astronomia, via Sansone 1, Sesto Fiorentino, 50019 Italy}
\affiliation[b]{INFN, Sezione di Firenze, via Sansone 1, Sesto Fiorentino, 50019 Italy}
\affiliation[c]{Universit{\'e}~de~Strasbourg,~CNRS,~IPHC~UMR~7178,~F-67000~Strasbourg,~France}
\affiliation[d]{Khalifa University of Science and Technology, Department of Physics, PO Box 127788, Abu Dhabi,   United Arab Emirates}
\affiliation[e]{IFIC - Instituto de F{\'\i}sica Corpuscular (CSIC - Universitat de Val{\`e}ncia), c/Catedr{\'a}tico Jos{\'e} Beltr{\'a}n, 2, 46980 Paterna, Valencia, Spain}
\affiliation[f]{INFN, Sezione di Napoli, Complesso Universitario di Monte S. Angelo, Via Cintia ed. G, Napoli, 80126 Italy}
\affiliation[g]{Universit{\`a} di Napoli ``Federico II'', Dip. Scienze Fisiche ``E. Pancini'', Complesso Universitario di Monte S. Angelo, Via Cintia ed. G, Napoli, 80126 Italy}
\affiliation[h]{INFN, Sezione di Roma, Piazzale Aldo Moro, 2 - c/o Dipartimento di Fisica, Edificio, G.Marconi, Roma, 00185 Italy}
\affiliation[i]{Universitat Polit{\`e}cnica de Catalunya, Laboratori d'Aplicacions Bioac{\'u}stiques, Centre Tecnol{\`o}gic de Vilanova i la Geltr{\'u}, Avda. Rambla Exposici{\'o}, s/n, Vilanova i la Geltr{\'u}, 08800 Spain}
\affiliation[j]{Subatech, IMT Atlantique, IN2P3-CNRS, Nantes Universit{\'e}, 4 rue Alfred Kastler - La Chantrerie, Nantes, BP 20722 44307 France}
\affiliation[k]{Universitat Polit{\`e}cnica de Val{\`e}ncia, Instituto de Investigaci{\'o}n para la Gesti{\'o}n Integrada de las Zonas Costeras, C/ Paranimf, 1, Gandia, 46730 Spain}
\affiliation[l]{Universit{\'e} Paris Cit{\'e}, CNRS, Astroparticule et Cosmologie, F-75013 Paris, France}
\affiliation[m]{Universit{\`a} di Genova, Via Dodecaneso 33, Genova, 16146 Italy}
\affiliation[n]{INFN, Sezione di Genova, Via Dodecaneso 33, Genova, 16146 Italy}
\affiliation[o]{LPC CAEN, Normandie Univ, ENSICAEN, UNICAEN, CNRS/IN2P3, 6 boulevard Mar{\'e}chal Juin, Caen, 14050 France}
\affiliation[p]{North-West University, Centre for Space Research, Private Bag X6001, Potchefstroom, 2520 South Africa}
\affiliation[q]{INFN, Sezione di Bologna, v.le C. Berti-Pichat, 6/2, Bologna, 40127 Italy}
\affiliation[r]{Universit{\`a} di Bologna, Dipartimento di Fisica e Astronomia, v.le C. Berti-Pichat, 6/2, Bologna, 40127 Italy}
\affiliation[s]{Universit{\`a} degli Studi della Campania "Luigi Vanvitelli", Dipartimento di Matematica e Fisica, viale Lincoln 5, Caserta, 81100 Italy}
\affiliation[t]{E.\,A.~Milne Centre for Astrophysics, University~of~Hull, Hull, HU6 7RX, United Kingdom}
\affiliation[u]{Czech Technical University in Prague, Institute of Experimental and Applied Physics, Husova 240/5, Prague, 110 00 Czech Republic}
\affiliation[v]{Comenius University in Bratislava, Department of Nuclear Physics and Biophysics, Mlynska dolina F1, Bratislava, 842 48 Slovak Republic}
\affiliation[w]{Nikhef, National Institute for Subatomic Physics, PO Box 41882, Amsterdam, 1009 DB Netherlands}
\affiliation[x]{Aix~Marseille~Univ,~CNRS/IN2P3,~CPPM,~Marseille,~France}
\affiliation[y]{INFN, Laboratori Nazionali del Sud, (LNS) Via S. Sofia 62, Catania, 95123 Italy}
% ...
\affiliation[be]{Universit{\'e} de Haute Alsace, rue des Fr{\`e}res Lumi{\`e}re, 68093 Mulhouse Cedex, France}
% ...





\abstract{The KM3NeT collaboration has reported the detection of the highest energy neutrino event observed to date. The energy of the event is of the order of $220\, \rm PeV$ hinting towards a  neutrino flux at the highest energies.
In this article, the potential blazar origin for this event is explored. The publicly available Astro-Multimessenger Modeling software is used to model the blazar gamma-ray and neutrino fluxes. It is concluded that a population of blazars could produce the diffuse flux compatible with the observation of the ultra-high energy event detected by the KM3NeT/ARCA detector. At the same time, the gamma-ray flux produced by such a population of blazars is consistent with the diffuse gamma-ray flux measured by the Fermi Large Area Telescope.} 

\maketitle
\flushbottom


\section{Introduction}
\label{sec:intro}
The KM3NeT collaboration is constructing two deep-sea neutrino detectors in the Mediterranean Sea: ARCA, optimised for the detection of high-energy cosmic neutrinos, offshore Sicily at a depth of about 3500 meters, and ORCA, off the coast of Toulon at about 2500 meters depth,  dedicated to neutrino oscillation studies. Both detectors consist of vertical detection units equipped with optical modules housing photomultiplier tubes~\cite{KM3NeT:2022pnv,KM3NeT:2025qao}. Although still under construction, both detectors are already operational and delivering first results, demonstrating their potential in the area of multi-messenger astrophysics.

On February 13, 2023, an ultra-high energy neutrino event was observed by the KM3NeT/ARCA detector, which was operating with 21 detection units. The event represents the most energetic neutrino detected to date, with a reconstructed energy of $220^{+570}_{-110}\,\mathrm{PeV}$ \cite{KM3NeT:2025npi}, which likely has an astrophysical origin. Soon after the publication of the discovery, several hypotheses have been suggested to interpret this event in an astrophysical framework. The possibility that the event is the signature of cosmogenic neutrinos has been explored in ref.~\cite{KM3NeT:2025vut}, high-energy neutrinos produced in the interaction of Ultra-High Energy Cosmic Rays (UHECRs) with the photons from the extragalactic background light and the cosmic microwave background. Although cosmogenic neutrinos are guaranteed given the UHECR spectra measured by the Pierre Auger~\cite{PierreAuger:2021hun} and the Telescope Array~\cite{Abu_Zayyad_2013} observatories, 
the large neutrino flux needed to account for KM3-230213A would imply that UHECRs should be injected up to very high redshift~$(z \simeq 6)$ with a subdominant proton component at the highest energies~\cite{KM3NeT:2025vut}. 
The possibility that the event is due to a powerful Galactic accelerator is discussed in ref.~\cite{KM3NeT:2025aps}. However, the lack of a gamma-ray counterpart as well as the need for this cosmic-ray accelerator to reach energies up to $\sim 4\cdot 10^{18}\, \rm eV$ make this hypothesis very unlikely. The possibility that the event might be produced by a single point-source is studied in ref.~\cite{KM3NeT:2025bxl}. Several blazars, Active Galactic Nuclei (AGNi) with relativistic jets oriented towards the Earth, are consistent with the position of the event~\cite{Urry:1995mg}. However, at present, the event is not associated with a specific source. 
Nonetheless, blazars remain well-motivated astrophysical sources~(see \cite{Giommi:2021bar} for a review). The IceCube Neutrino Observatory has found a correlation of $\sim 3\sigma$ for high-energy neutrino emission with a blazar named TXS 0506+056 ~\cite{IceCube:2018dnn,IceCube:2022der} confirming that blazar jets are accelerators of cosmic rays up to at least $\sim 1-10\, \rm PeV$ energies.
Also, the ANTARES collaboration reported a $\sim 2\sigma$ hint of correlation between neutrino events and radio flaring blazars~\cite{ANTARES:2023lck}. Furthermore, recent phenomenological analyses~\cite{Rodrigues:2024fhu} have shown that blazars might emit neutrinos in the energy range between $\sim 1-100\, \rm PeV$.

In this article, the possibility that blazars accelerate cosmic rays along their jet axis and emit ultra-high energy neutrinos is studied. 
The complexity of blazar emissions is discussed under the most conservative astrophysical assumptions, so that the conclusions of this study are not influenced by model-specific details.
High-energy gamma-ray and neutrino emissions from blazars are modelled using the publicly available Astro-Multimessenger Modeling~(AM3) software~\cite{Klinger:2023zzv}. This code implements a generic one-zone model for the relativistic jets and incorporates particle acceleration and emission processes, including both leptonic and hadronic interactions. A statistical analysis is performed to probe the parameter space and investigate under which condition blazars might have produced the event. In this regard, any diffuse flux interpretation of the event must be compatible with the diffuse gamma-ray measurements from the Fermi Large Area Telescope (Fermi-LAT) and the lack of a counterpart from other neutrino telescopes such as IceCube and Auger~\cite{Fermi-LAT:2014ryh,IceCube:2025ezc,PierreAuger:2019ens,AbdulHalim:2023SN,PierreAuger:2025jwt}.
 This analysis shows that a blazar population is consistent with observational constraints and can produce the required flux for KM3-230213A.
 
The article is organised as follows: the single blazar geometrical setup and the model assumptions for the simulation of the  neutrino and gamma-ray emissions are described in Sec.~\ref{Sec:Single_Blazar_model}. The diffuse flux calculation is discussed in Sec.~\ref{Sec:Diffuse_flux_calculation}, and the statistical analysis used to process the different datasets is presented in Sec.~\ref{Sec:Statistical_analysis}. The results and conclusions of this work are reported in Secs.~\ref{Sec:results_discussion} and \ref{Sec:conclusions}. Finally, a statistical analysis using KM3NeT/ARCA-only information to show that the result is compatible with the gamma-ray constraints imposed by the Fermi-LAT collaboration is presented in Appendix~\ref{app:only_km3}.

\section{Blazar simulation: source model and assumptions}\label{Sec:Single_Blazar_model}

In this section, details of the simulation of high-energy neutrino emitters from a single blazar are discussed. The AM3 software, which solves time-dependent transport equations for primary and secondary particles in astrophysical jets, is used to predict multi-messenger emissions. 
It implements a single-zone model where the emission region is treated as a compact spherical blob moving with relativistic velocity along the jet. 

The jet is modelled as a relativistic outflow with a bulk Lorentz factor \( \Gamma_b \) fixed at 17.6, the mean value obtained by~\cite{Rodrigues:2023vbv} (also adopted as the default in the AM3~software~\cite{Klinger:2023zzv}), which defines the jet speed relative to the observer. Following \cite{Rodrigues:2023vbv}, the jet is observed at an angle \( \theta_{\text{obs}} = \frac{1}{\Gamma_b} \) relative to its axis, which results in a Doppler factor \( \delta_D = \Gamma_b \), contributing to Doppler boosting of the observed radiation.

The particle acceleration is described by a power-law with spectral indices $\alpha_p$ and $\alpha_e$ for protons and electrons, respectively.
In the following analysis, the maximum electron and proton energies, in the jet reference frame,  are fixed at $25\, \rm GeV$ and $30\, \rm PeV$, respectively. These values are consistent with the average value determined by using a catalogue of 32 blazars~\cite{Rodrigues:2024fhu}. 
Larger values of the maximum proton energy do not affect the conclusions of this article, while smaller ones would prevent blazars from producing neutrinos energetic enough to explain KM3-230213A.

The other parameters defined in the AM3 simulations are:

\begin{itemize}
    \item Radius~$(R^{'})$: characteristic size of the blazar’s emission region.
    \item Magnetic Field (\( B' \)): strength of the magnetic field within the jet.

 \item  $R_{\rm{BLR}}$: the radius of the Broad Line Region,  fast-moving gas clouds responsible for broad optical and UV lines.
    
      \item  $R_{\rm diss}$: dissipation radius in the jet, relative to R$_{BLR}$ controlling external photon impact.
 
      \item $\Gamma_{\text{e~max}}$: Maximum  Lorentz factor for electrons.
     \item $\Gamma_{\text{p~max}}$: Maximum  Lorentz factor for protons.
  
    
     \item Electron Luminosity (\( L_e \)): luminosity associated with the relativistic electrons in the jet.
      
    \item Proton Luminosity (\( L_p \)): luminosity associated with the relativistic protons in the jet.
    \item Baryonic Loading ($\eta = L_p/L_e$): this parameter characterises the ratio of proton to electron luminosity.
\end{itemize}

Since it is not feasible to account for the variability of each parameter across the whole blazar population, some of the above-mentioned parameters are fixed at the average values reported in ~\cite{Rodrigues:2024fhu},
which are consistent with the parameter ranges derived from the catalogue of 324 nearby blazars in \cite{Rodrigues:2023vbv}. These values are summarised in Table~1 and define the baseline AM3 source template.

\begin{table}[h!]
    \centering
    \renewcommand{\arraystretch}{1.2}
    \begin{tabular}{|c|c|c|c|c|c|c|c|c|c|}
        \hline
        \begin{tabular}{@{}c@{}}$ \log_{10}(R') $ \\ $[\mathrm{cm}]$\end{tabular} &
        \begin{tabular}{@{}c@{}}$ \Gamma_b $ \\ $ $\end{tabular} &
        \begin{tabular}{@{}c@{}}$ B' $ \\ $[\mathrm{G}]$\end{tabular} &
        \begin{tabular}{@{}c@{}}$ \log_{10}(R_{\text{diss}}) $ \\ $[\mathrm{cm}]$\end{tabular} &
        \begin{tabular}{@{}c@{}}$ \frac{R_{\text{diss}}}{R_{\text{BLR}}} $ \\ $  $\end{tabular} &
        \begin{tabular}{@{}c@{}}$ \log_{10}(\Gamma_{\text{e~max}}) $ \\ $ $\end{tabular} &
        \begin{tabular}{@{}c@{}}$ \log_{10}(\Gamma_{\text{p~max}}) $ \\ $ $\end{tabular} &
        \begin{tabular}{@{}c@{}}$ \alpha_e $ \\ $ $\end{tabular} &
        \begin{tabular}{@{}c@{}}$ \log_{10}(L_{\rm e}) $ \\ $[\mathrm{erg/s}]$\end{tabular} \\
        \hline
        15.9 & 17.6 & 2.6 & 16.8 & 1.8 & 4.7 & 7.5 & 1.7 & 41.2 \\
        \hline
    \end{tabular}
    \caption{Fixed source parameters used in the AM3 simulation. Values correspond to the average reported in ~\cite{Rodrigues:2024fhu}.
    }
    \label{tab:source_parameters}
\end{table}


In the following, the baryonic loading~$(\eta)$, and the proton spectral index~$(\alpha_p)$ are treated as free parameters. 
The baryonic loading modifies the normalisation of the gamma-ray and neutrino spectra while the spectral index determines their shape. These two parameters effectively capture the full variability of these spectra among the sources.
\vspace{3pt}

The AM3 software implements all the main processes for the production of secondary particles. They are summarised in the following.

\begin{itemize}
    \item Synchrotron Radiation: \\Electrons accelerated in the jet produce synchrotron radiation.\\
    \item Inverse Compton Scattering~(ICS): \\Both synchrotron self-Compton and external Compton  scattering contribute to the observed high-energy radiation from the jet:\\
     \hspace*{3.5cm} $\quad e^{-} + \gamma \rightarrow e^{-} + \gamma$.

    \item Bethe-Heitler Pair Production: \\Protons interacting with photons in the jet produce electron-positron pairs:\\
     \hspace*{3.5cm} $ \quad p + \gamma \rightarrow p + e^{+} + e^{-} $.
    
    \item Pair Production:\\ Photon-photon interactions lead to the creation of electron-positron pairs:\\
     \hspace*{3.5cm} $\quad \gamma + \gamma \rightarrow e^{+} + e^{-}$.
    \item Pion Decay from photomeson interactions:\\ High-energy protons in the jet interact with ambient photons, producing pions that decay into gamma-rays and neutrinos:\\
    \hspace*{3.5cm}  $\quad p + \gamma  \rightarrow \begin{cases} p  + \pi^{0} \\ n + \pi^{+}  \end{cases} \rightarrow \begin{cases} 2\gamma + X \\ \nu_{\mu} + e^{+} + \nu_{e} + \bar{\nu}_{\mu} + X \end{cases}$.
\end{itemize}

\begin{figure}[h!]
    \centering
    \includegraphics[width=\linewidth]{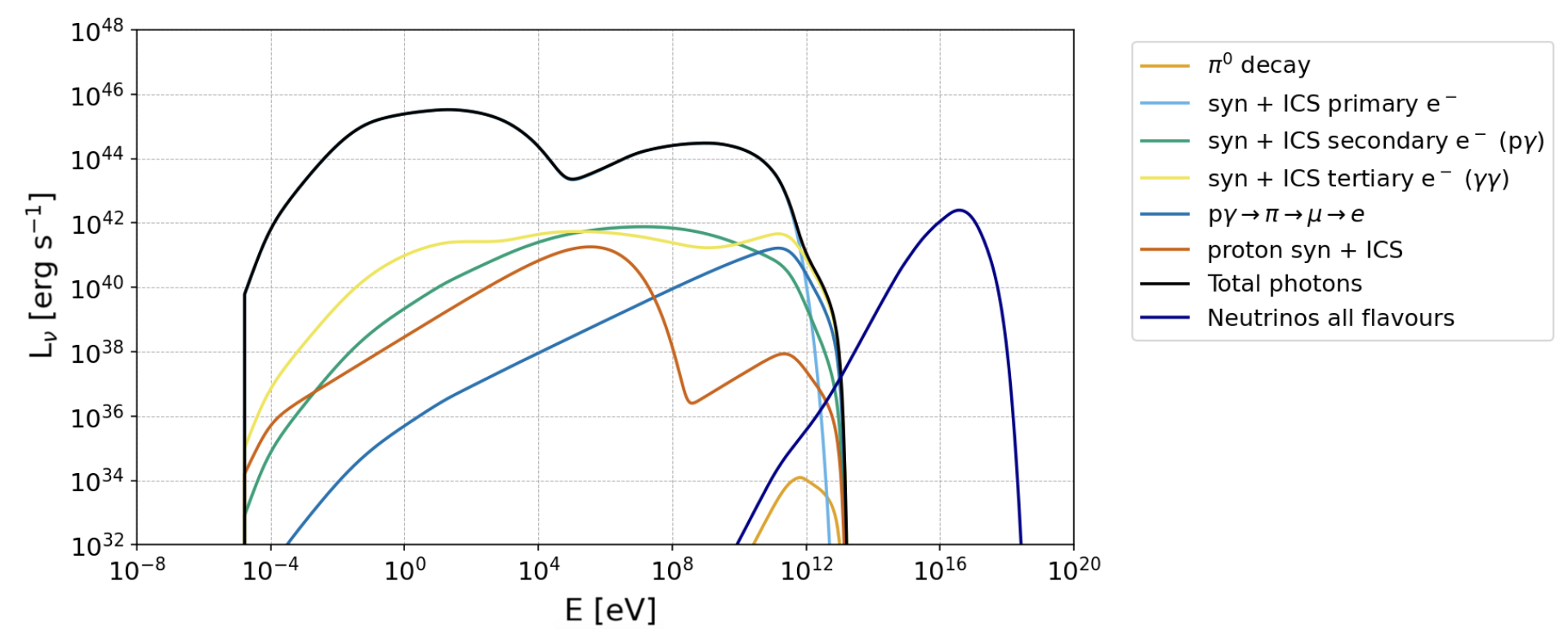}
    \caption{Differential luminosity for gamma-rays and neutrinos (all-flavour) as a function of the energy. The values of the baryonic loading and the proton spectral index have been fixed at $10$ and $1.8$, respectively.}
    \label{fig:differential_luminosity_benchmark}
\end{figure}

Modelling these processes using AM3 allows for the computation of the differential photon flux and the differential neutrino flux from a single simulated blazar. A benchmark case of the luminosity~(in unit of $\rm erg\, \rm s^{-1}$) is shown in Fig.~\ref{fig:differential_luminosity_benchmark} for a single blazar as a function of the energy, using $\eta = 10$ and $\alpha_p = 1.8$ (these values correspond to the best-fit values found by this analysis, see below for more details). For completeness, the different processes are shown separately to evaluate their individual contribution to the overall gamma-ray spectrum and to the overall neutrino spectrum. 

\section{Diffuse flux calculation}\label{Sec:Diffuse_flux_calculation}

The blazar diffuse flux is calculated through the gamma-ray luminosity function, following~\cite{Ajello:2015mfa}.  The luminosity function at $z=0$  is defined as: 
\begin{equation}
    \mathcal{S}(L_\gamma, z=0, \Gamma) = \frac{dN}{dL_\gamma~dV_c~d\Gamma} = \frac{A}{\ln(10)~L_\gamma} \left( \left(\frac{L_\gamma}{L_*}\right)^{\gamma_1} + \left(\frac{L_\gamma}{L_*}\right)^{\gamma_2} \right)^{-1} \times e^{-0.5 \frac{[\Gamma_{\rm ph} - \mu(L_\gamma)]^2}{\sigma^2}},
\end{equation}
\noindent
where $L_{\gamma}$ is the $0.1-100\, \rm GeV$ rest-frame gamma-ray integrated luminosity in units of $\rm GeV\, \rm s^{-1}$ and $V_c$ is the comoving volume. $\Gamma_{\rm ph} $ represents the photon flux index that follows a Gaussian distribution, with  $\mu$ and $\sigma$ denoting the mean and dispersion of the distribution, respectively. $\mu$ is expressed in terms of $L_\gamma$ as:
\begin{equation}
\mu(L_\gamma) = \mu_* + \beta \times \bigg[\log_{10}(L_\gamma) - 46\bigg].
\end{equation}
The luminosity function at a given redshift $z$ is then given by:
\begin{equation}
    \mathcal{S}(L_\gamma, z, \Gamma) = \mathcal{S}(L_\gamma, z=0, \Gamma) \times e(z, L_\gamma),
\end{equation}
where
\begin{equation}
    e(z, L_\gamma) = (1+z)^{k_d(L_{\gamma})} e^{\frac{z}{\zeta}}, 
\end{equation}
with $k_d(L_{\gamma}) = k_{*} + \tau_{\gamma} (\rm log_{10} (L_{\gamma}) -46)$. 
\\
All the numerical values of the parameters  $(A, L_{*}, \gamma_1, \gamma_2, k_{*}, \zeta, \tau_{\gamma}, \mu_{*}, \beta, \sigma)$ are reported in Tab.~\ref{tab:parameter_luminosity_function}.
Since in the model used in this analysis there is no $\Gamma_{\rm ph}$ parameter, the final luminosity function~$\mathcal{S}(L_{\gamma},z)$ is given by

\begin{equation}
    \mathcal{S}(L_{\gamma},z) = \int_{1}^{3.5} d\Gamma_{\rm ph} \,  \mathcal{S}(L_{\gamma},z,\Gamma_{\rm ph}).
\end{equation}
\noindent
This is the same approach used in ref.~\cite{Palladino:2018lov}. The diffuse blazar flux is calculated with

\begin{align}\label{eq:diff_flux}
\Phi_{\rm Diff}^{\nu, \gamma}(E,\eta, \alpha_p)
&= \int_{10^{-3}}^{6} dz 
   \int_{10^{43}\, \rm erg\, s^{-1}}^{10^{52}\, \rm erg\, s^{-1}} 
   dL_{\gamma}\, \mathcal{S}(L_{\gamma}, z) \nonumber\\
&\quad \times 
   \frac{dN}{dE\, dA\, dt}(E, \eta, \alpha_p, L_{\gamma}, z)\,
   e^{-\tau^{\nu, \gamma}(E,z)}\,
   \frac{dV_c}{dz}(z) \, ,
\end{align}
\noindent
where $\frac{dN}{dE dA dt} (E, \eta, \alpha_p, L_{\gamma},z)$  describes the single blazar gamma-ray and neutrino spectra from AM3 including the correction for energy redshift. $\frac{dV_c}{dz}(z)$ represents the Jacobian of the transformation from~$V_c$ to~$z$. $\tau^{\nu}(E,z)$ is the neutrino optical depth equal to zero because they propagate unimpeded, and $\tau^{\gamma}(E,z)$ is the gamma-ray optical depth which is implemented following ref.~\cite{2011MNRAS.410.2556D}.

\begin{table}[h!]
    \centering
    \renewcommand{\arraystretch}{1.2}
    \begin{tabular}{|c|c|c|c|c|c|c|c|c|c|}
        \hline
        \begin{tabular}{@{}c@{}}$ A $ \\ $[\mathrm{Gpc^{-3}}]$ \end{tabular} &
        \begin{tabular}{@{}c@{}}$ \rm log_{10} (L_{*}) $ \\ $[\rm{erg\,s^{-1}}]$ \end{tabular} &
        \begin{tabular}{@{}c@{}}$ \gamma_1 $ \\ $ $ \end{tabular} &
        \begin{tabular}{@{}c@{}}$ \gamma_2 $ \\ $ $ \end{tabular} &
        \begin{tabular}{@{}c@{}}$ k_{*} $ \\ $ $ \end{tabular} &
        \begin{tabular}{@{}c@{}}$ \zeta $ \\ $ $ \end{tabular} &
        \begin{tabular}{@{}c@{}}$ \tau_{\gamma} $ \\ $ $ \end{tabular} &
        \begin{tabular}{@{}c@{}}$ \mu_{*} $ \\ $ $ \end{tabular} &
        \begin{tabular}{@{}c@{}}$ \beta $ \\ $ $ \end{tabular} &
        \begin{tabular}{@{}c@{}}$ \sigma $ \\ $ $ \end{tabular} \\
        \hline
        $1.22\cdot 10^{-2}$ & $47.64$ & $2.80$ & $1.26$ & $12.14$ & $-0.15$ & $2.79$ & $2.22$ & $0.10$ & $0.28$ \\
        \hline
    \end{tabular}
    \caption{Values of the parameters for the luminosity distribution function~\cite{Ajello:2015mfa}.}
    \label{tab:parameter_luminosity_function}
\end{table}

\section{Likelihood analysis}\label{Sec:Statistical_analysis}

In order to interpret the KM3-230213A event in terms of a diffuse neutrino flux from blazars, the following constraints must be satisfied: 
\begin{itemize}
    \item the diffuse neutrino flux from blazars should be compatible with the non-observation of high-energy neutrino events reported by the IceCube and Auger collaborations;
    \item the diffuse gamma-ray flux from blazars cannot be larger than the extragalactic background  gamma-ray flux measured by the Fermi-LAT collaboration, also known as the EGB spectrum~\cite{Fermi-LAT:2014ryh}. 
\end{itemize}

For the neutrino part, the likelihood function describing the neutrino emission is defined as the Poisson probability of observing $n_s = 1$ event with an expected  number of events given by $\lambda^{\rm tot}(\eta, \alpha_p)$
\begin{equation}\label{eq:lik_nu}
    \mathcal{L}_{\nu}(\eta, \alpha_p) = \lambda^{\rm tot}(\eta, \alpha_p) e^{-\lambda^{\rm tot}(\eta, \alpha_p)},
\end{equation}
where $\lambda^{\rm tot}(\eta, \alpha_p)$ is estimated as 
\begin{equation}
    \lambda^{\rm tot} (\eta, \alpha_p) =  \int_{\Delta E} \epsilon^{\rm tot}_{\rm eff} (E)  \Phi_{\rm Diff}^{\nu}(E,\eta, \alpha_p)dE,
\end{equation}
where $\epsilon^{\rm tot}_{\rm eff}(E) = 4\pi (T_{\rm KM3} A^{\rm KM3}_{\rm eff}(E)  + T_{\rm IC} A^{\rm IC}_{\rm eff}(E))$ is the sum of the KM3NeT~\cite{KM3NeT:2025npi} and IceCube~\cite{IceCube:2025ezc} exposures. $T_{\rm KM3} = 335\,  \rm days$ and $T_{\rm IC} = 12.6 \, \rm yr$ are the KM3NeT and IceCube lifetimes, respectively. As in~\cite{KM3NeT:2025ccp}, measurements from both detectors are combined for energies above $10\, \rm PeV$. In this work, the exposure of Pierre Auger~\cite{PierreAuger:2019ens} is not included, because IceCube currently provides higher sensitivity in the energy range where blazars are assumed to emit neutrinos. In order to suppress the parameters producing a diffuse gamma-ray flux overshooting the Fermi-LAT measurements, a Gaussian penalty term is introduced in the likelihood function as in ref.~\cite{Ambrosone:2020evo}: 

\begin{equation}\label{eq:lik_gamma}
    \mathcal{L}_{\gamma}(\eta, \alpha_p) = e^{-\frac{1}{2}~\big(\frac{f(\eta, \alpha_p) - 0.86 }{ 0.14}\big)^2},
\end{equation}
where $f$ represents the integrated fraction of the diffuse gamma-ray spectrum between $50-2000\, \rm GeV$:
\begin{equation}\label{eq:blazarfraction}
    f(\eta, \alpha_p) = \frac{\int_{50\, \rm GeV}^{2000\, \rm GeV} dE \,  \Phi_{\rm diff}^{\gamma}(E,\eta, \alpha_p) }{F_{\rm {Fermi- LAT}}^{\gamma} },
\end{equation}
and $ F_{\rm {Fermi-LAT}}^{\gamma} = 2.4\cdot 10^{-9} \rm photon \,  \rm cm^{-2}\, \rm s^{-1}\, \rm sr^{-1}$ is the total diffuse flux measured by the Fermi-LAT telescope~\cite{Fermi-LAT:2015otn}. 
This fraction has been estimated in ref.~\cite{Fermi-LAT:2015otn} to be $ (86\%^{+16\%}_{-14\%})$. Although the value ascribed to blazars is subject to significant uncertainties due to specific assumptions~\cite{Lisanti:2016jub}, in this analysis this fraction is used solely to avoid exceeding the Fermi-LAT measurements. Therefore, this uncertainty does not affect the final conclusions of the present work. The total likelihood of the model is the product of the photon and neutrino likelihood functions:
\begin{equation}
    \mathcal{L}_{\rm tot}(\eta, \alpha_p) = \mathcal{L}_{\nu}(\eta, \alpha_p) \mathcal{L}_{\gamma}(\eta, \alpha_p).
\end{equation}

The parameter space is explored using the test statistic (TS) defined as:
\begin{equation}
    \rm TS(\eta, \alpha_p) = 2 \cdot \rm ln \bigg(\frac{\mathcal{L}_{\rm tot}(\Tilde{\eta},\Tilde{\alpha_p})}{\mathcal{L}_{\rm tot}(\eta, \alpha_p)}\bigg),
\end{equation}
where  $\Tilde{\eta} $ and $\Tilde{\alpha_p}$ are the parameters that maximise the overall likelihood. According to Wilks' theorem~\cite{Wilks:1938dza}, the test statistic follows a chi-squared distribution with 2 degrees of freedom, as the number of free parameters.

\section{Results and Discussion}\label{Sec:results_discussion}

\begin{figure}[h!]
    \centering
     \includegraphics[width=\linewidth]{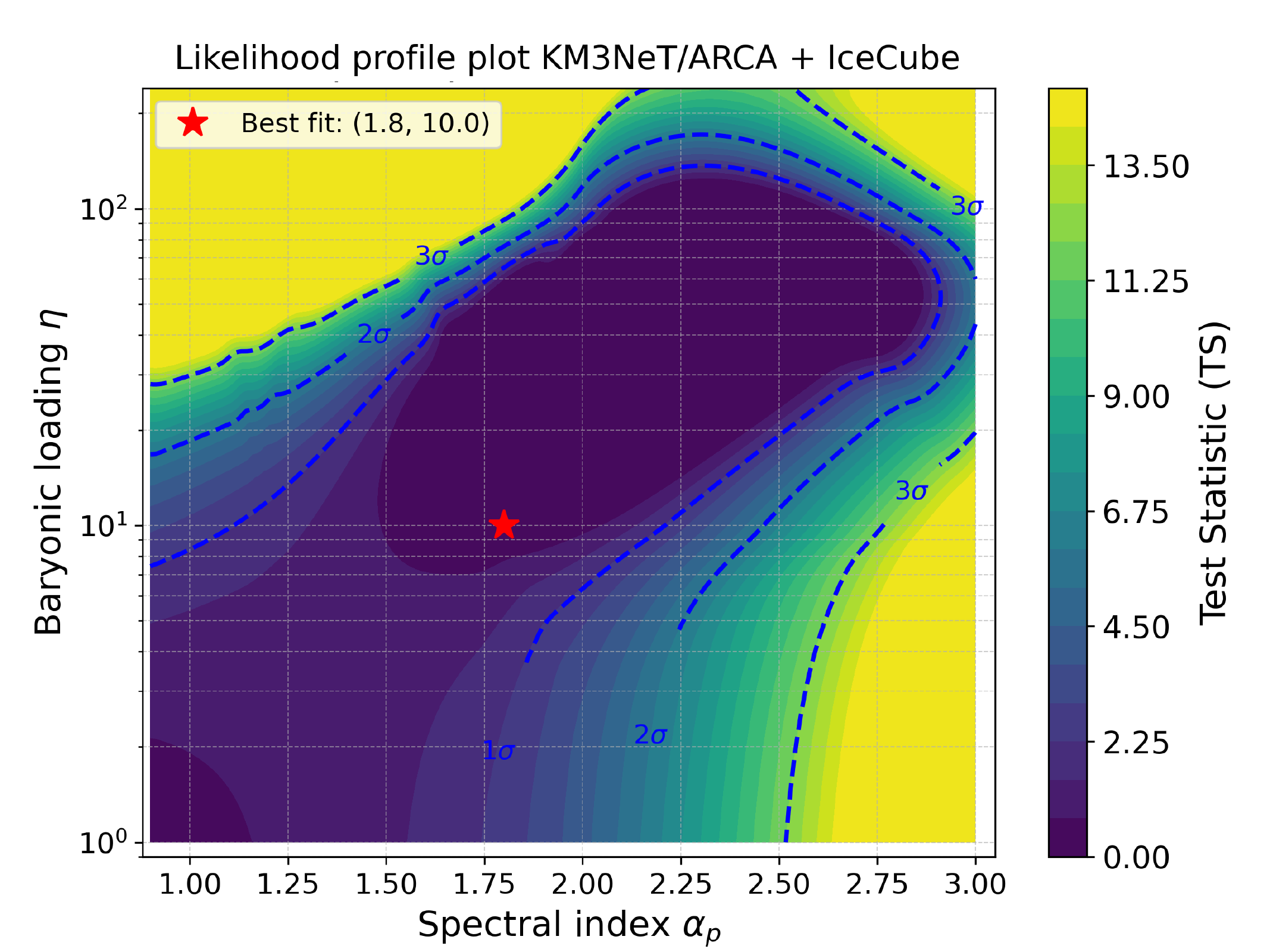}
    \caption{Test statistic contour plot in terms of baryonic loading~($\eta$) and the proton spectral index~($\alpha_p$) for the joint KM3NeT/ARCA and IceCube analysis. The best-fit value is reported with a red star.}
    \label{bestfitkm3icecube}
\end{figure}

The $1,2, 3 \,  \sigma$ TS contours are shown in Figure \ref{bestfitkm3icecube}. The best fit indicates values of $\Tilde{\eta} \approx 10$ and $\Tilde{\alpha_p} \approx 1.8$.
The result also suggests that $\eta$ values larger than $100$ are strongly disfavoured above the $3\sigma $ level. Furthermore, the analysis favours hard spectral indices, $\alpha_p \lesssim 2$, which maximise the expected number of events in KM3NeT/ARCA and IceCube within the considered energy range $\Delta E$.

\begin{figure}[h!]
    \centering    
    \includegraphics[width=\linewidth]{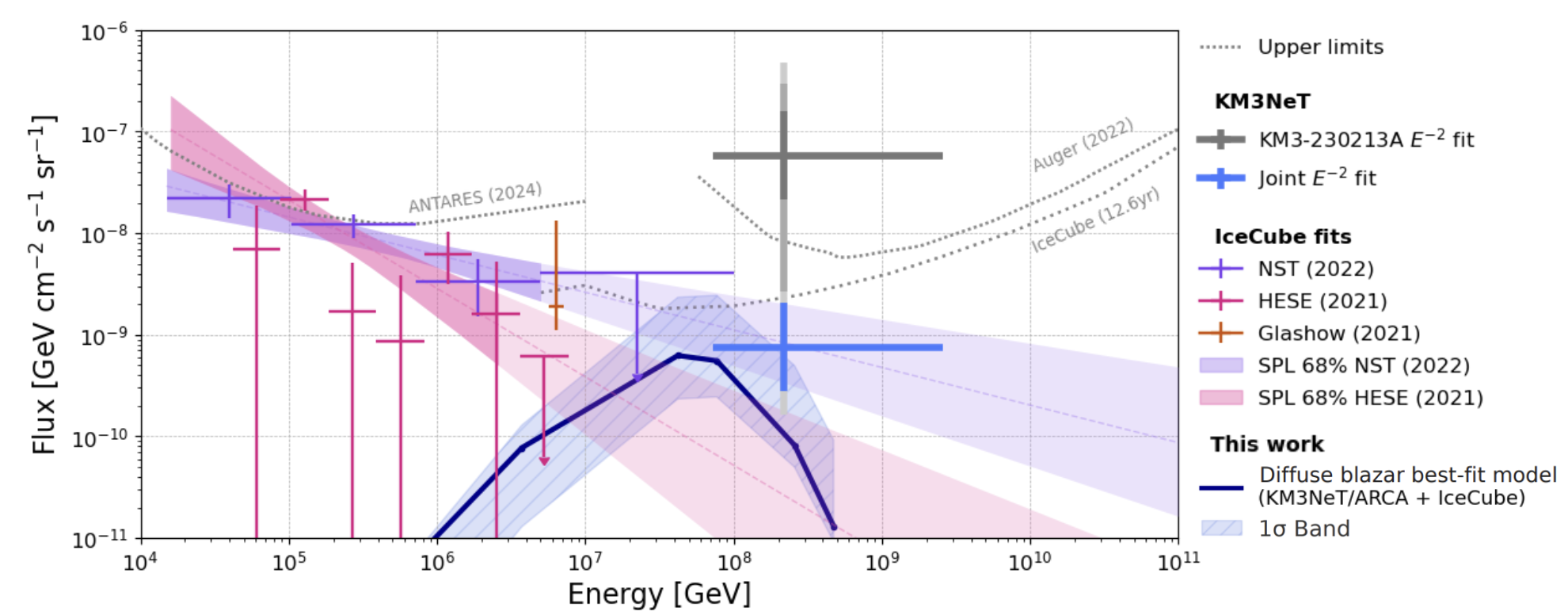}
    \caption{Neutrino  diffuse spectral energy distribution  for blazars in terms of the energy for a single neutrino flavour. The dark blue solid line represents the best fit, while the shaded region is the $1\sigma$ band. The prediction is compared with the KM3-230213A equivalent flux~\cite{KM3NeT:2025npi}, the joint $E^{-2}$ flux obtained by~\cite{KM3NeT:2025ccp} including IceCube-Extreme High-Energy~\cite{IceCube:2018fhm} and Auger non-observations, and the updated IceCube~\cite{IceCube:2025ezc} and Auger differential upper limits~\cite{AbdulHalim:2023SN}. For comparison, the diffuse neutrino flux measured by the IceCube Neutrino Observatory with several samples~\cite{IceCube:2020wum,Abbasi:2021qfz,IceCube:2021rpz} and also the ANTARES upper limits~\cite{ANTARES:2024ihw} are reported. The pink and purple shaded regions represent the IceCube single-power-law (SPL) fits for  High-Energy Starting Events~(HESE)~\cite{IceCube:2020wum} and Northern Sky Tracks~(NST)~\cite{Abbasi:2021qfz}, respectively.} %

    \label{final_km3_icecube}
\end{figure}

\begin{figure}[h!]
    \centering        
    \includegraphics[width=\linewidth]{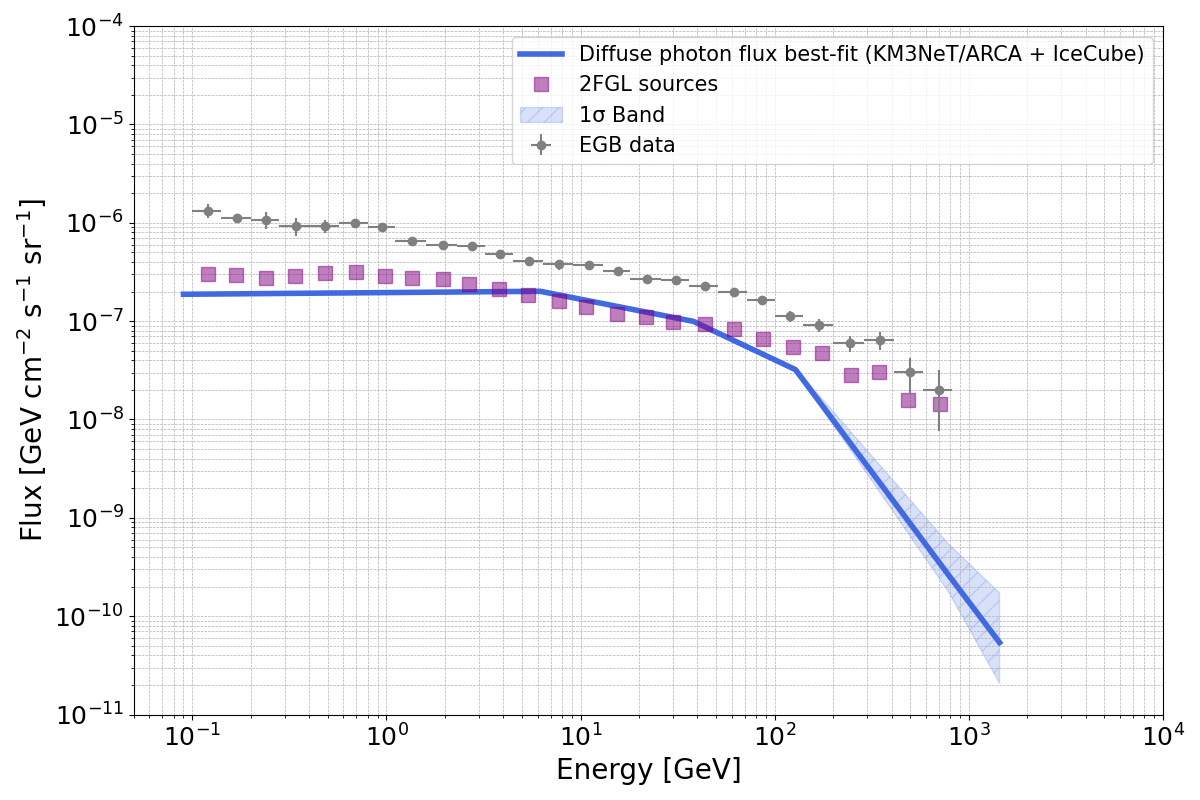}
    \caption{Gamma-ray diffuse spectral energy distribution  for blazars as a function of the energy. The solid line represents the best fit while the shaded region the $1\sigma$ band. The result is compared with ExtraGalactic Background  measurements of Fermi-LAT and the 2FGL sources~\cite{Fermi-LAT:2014ryh}.}
    \label{fig:photon_diffuse_flux}
\end{figure}
The statistical analysis is dominated by the neutrino contribution to the likelihood~(Eq.~\ref{eq:lik_nu})  and yields a predicted neutrino spectrum, shown in Fig.~\ref{final_km3_icecube}, which closely matches the joint $E^{-2}$ fit reported in ref.~\cite{KM3NeT:2025ccp}.
The $1\sigma$ band for the blazar diffuse flux is compatible with the differential upper limits reported by IceCube~\cite{IceCube:2025ezc} and Auger~\cite{AbdulHalim:2023SN}. The spectrum is also consistent with the diffuse neutrino flux upper limits derived from other IceCube data samples~\cite{IceCube:2020wum,Abbasi:2021qfz,IceCube:2021rpz} and from the ANTARES collaboration~\cite{ANTARES:2024ihw}. In this scenario, blazars contribute negligibly to the diffuse neutrino flux at energies $\lesssim 1\, \rm PeV$.
The corresponding gamma-ray spectrum is compatible with the 2FGL resolved source catalogue~\cite{Fermi-LAT:2014ryh}, with blazars contributing $\simeq 42\%$  of the EGB as shown in Fig.~\ref{fig:photon_diffuse_flux}. Although this value is lower than the 86\% fraction adopted in Eq.~\ref{eq:lik_gamma}, this does not indicate any tension with the gamma-ray flux. In fact, the total diffuse gamma-ray flux is dominated by leptonic contributions~(see Fig.~\ref{fig:differential_luminosity_benchmark}), and this analysis cannot constrain all the parameters listed in Tab.~\ref{tab:source_parameters}, such as the electron luminosity. Therefore, the adopted blazar fraction~(Eq.~\ref{eq:blazarfraction}) is regarded as an upper limit to avoid overproducing gamma rays.

This suggests that blazars may contribute significantly to the
UHE neutrino emission leading to KM3-230213A. Notably, even when considering only KM3NeT data (see Appendix~\ref{app:only_km3} for details), the inferred diffuse neutrino spectrum remains consistent with EGB measurements. In this case, the most stringent constraints arise from diffuse neutrino observations in the $\sim 1\text{--}10\, \rm PeV$ range, where blazars nearly saturate the upper limits. To explain such a high neutrino flux, blazars would need to be approximately an hundred times more luminous, corresponding to baryonic loadings $ \eta \simeq 10^3$.

\section{Conclusions}\label{Sec:conclusions}



The KM3NeT collaboration has  reported the highest-energy neutrino event detected to date, with an estimated energy of $\sim 220\, \rm PeV$. In this article, the possibility that this event might have originated from the diffuse neutrino flux produced by blazars is considered. This is theoretically well motivated since blazars are the most luminous objects in the gamma-ray sky. In order to constrain the spectrum of a single blazar, the publicly available AM3 software is used. In order to extrapolate this result to the whole blazar population, the luminosity function inferred by the Fermi-LAT collaboration is used. The constraints imposed by the IceCube non-observation and  the gamma-ray  measurements from Fermi-LAT are also taken into account. Results demonstrate that  the diffuse neutrino flux due to blazars is compatible with all constraints and may explain the KM3-230213A event considering a joint fit combining information with the IceCube diffuse flux measurement. 
In Appendix~\ref{app:only_km3}, the case of the KM3NeT/ARCA exposure alone is discussed. In this case, the diffuse emission from blazars remains consistent with Fermi-LAT gamma-ray constraints, though it is in tension with the IceCube upper limits at the $\sim 2.5$–$3\,\sigma$ level.

\acknowledgments

The authors acknowledge the financial support of:
KM3NeT-INFRADEV2 project, funded by the European Union Horizon Europe Research and Innovation Programme under grant agreement No 101079679;
Funds for Scientific Research (FRS-FNRS), Francqui foundation, BAEF foundation.
Czech Science Foundation (GAČR 24-12702S);
Agence Nationale de la Recherche (contract ANR-15-CE31-0020), Centre National de la Recherche Scientifique (CNRS), Commission Europ\'eenne (FEDER fund and Marie Curie Program), LabEx UnivEarthS (ANR-10-LABX-0023 and ANR-18-IDEX-0001), Paris \^Ile-de-France Region, Normandy Region (Alpha, Blue-waves and Neptune), France,
The Provence-Alpes-Côte d'Azur Delegation for Research and Innovation (DRARI), the Provence-Alpes-Côte d'Azur region, the Bouches-du-Rhône Departmental Council, the Metropolis of Aix-Marseille Provence and the City of Marseille through the CPER 2021-2027 NEUMED project,
The CNRS Institut National de Physique Nucléaire et de Physique des Particules (IN2P3);
Shota Rustaveli National Science Foundation of Georgia (SRNSFG, FR-22-13708), Georgia;
This research was funded by the European Union (ERC MuSES project No 101142396); 
The General Secretariat of Research and Innovation (GSRI), Greece;
Istituto Nazionale di Fisica Nucleare (INFN) and Ministero dell’Universit{\`a} e della Ricerca (MUR), through PRIN 2022 program (Grant PANTHEON 2022E2J4RK, Next Generation EU) and PON R\&I program (Avviso n. 424 del 28 febbraio 2018, Progetto PACK-PIR01 00021), Italy; IDMAR project Po-Fesr Sicilian Region az. 1.5.1; A. De Benedittis, W. Idrissi Ibnsalih, M. Bendahman, A. Nayerhoda, G. Papalashvili, I. C. Rea, A. Simonelli have been supported by the Italian Ministero dell'Universit{\`a} e della Ricerca (MUR), Progetto CIR01 00021 (Avviso n. 2595 del 24 dicembre 2019); KM3NeT4RR MUR Project National Recovery and Resilience Plan (NRRP), Mission 4 Component 2 Investment 3.1, Funded by the European Union – NextGenerationEU,CUP I57G21000040001, Concession Decree MUR No. n. Prot. 123 del 21/06/2022;
Ministry of Higher Education, Scientific Research and Innovation, Morocco, and the Arab Fund for Economic and Social Development, Kuwait;
Nederlandse organisatie voor Wetenschappelijk Onderzoek (NWO), the Netherlands;
The grant “AstroCeNT: Particle Astrophysics Science and Technology Centre”, carried out within the International Research Agendas programme of the Foundation for Polish Science financed by the European Union under the European Regional Development Fund; The program: “Excellence initiative-research university” for the AGH University in Krakow; The ARTIQ project: UMO-2021/01/2/ST6/00004 and ARTIQ/0004/2021;
Ministry of Education and Scientific Research, Romania;
Slovak Research and Development Agency under Contract No. APVV-22-0413; Ministry of Education, Research, Development and Youth of the Slovak Republic;
MCIN for PID2021-124591NB-C41, -C42, -C43 and PDC2023-145913-I00 funded by MCIN/AEI/10.13039/501100011033 and by “ERDF A way of making Europe”, for ASFAE/2022/014 and ASFAE/2022 /023 with funding from the EU NextGenerationEU (PRTR-C17.I01) and Generalitat Valenciana, for Grant AST22\_6.2 with funding from Consejer\'{\i}a de Universidad, Investigaci\'on e Innovaci\'on and Gobierno de Espa\~na and European Union - NextGenerationEU, for CSIC-INFRA23013 and for CNS2023-144099, Generalitat Valenciana for CIDEGENT/2020/049, CIDEGENT/2021/23, CIDEIG/2023/20, ESGENT2024/24, CIPROM/2023/51, GRISOLIAP/2021/192 and INNVA1/2024/110 (IVACE+i), Spain;
Khalifa University internal grants (ESIG-2023-008, RIG-2023-070 and RIG-2024-047), United Arab Emirates;
The European Union's Horizon 2020 Research and Innovation Programme (ChETEC-INFRA - Project no. 101008324).

Views and opinions expressed are those of the author(s) only and do not necessarily reflect those of the European Union or the European Research Council. Neither the European Union nor the granting authority can be held responsible for them.

\bibliography{references}

\bibliographystyle{unsrtnat}

\newpage

\appendix

\section{Likelihood Analysis with Only KM3NeT Information}\label{app:only_km3}

In this section, the results of the likelihood analysis are reported, considering only the KM3NeT/ARCA exposure in Eq.~\ref{eq:lik_nu}. 

\begin{figure}[h!]
    \centering
    \includegraphics[width=\linewidth]{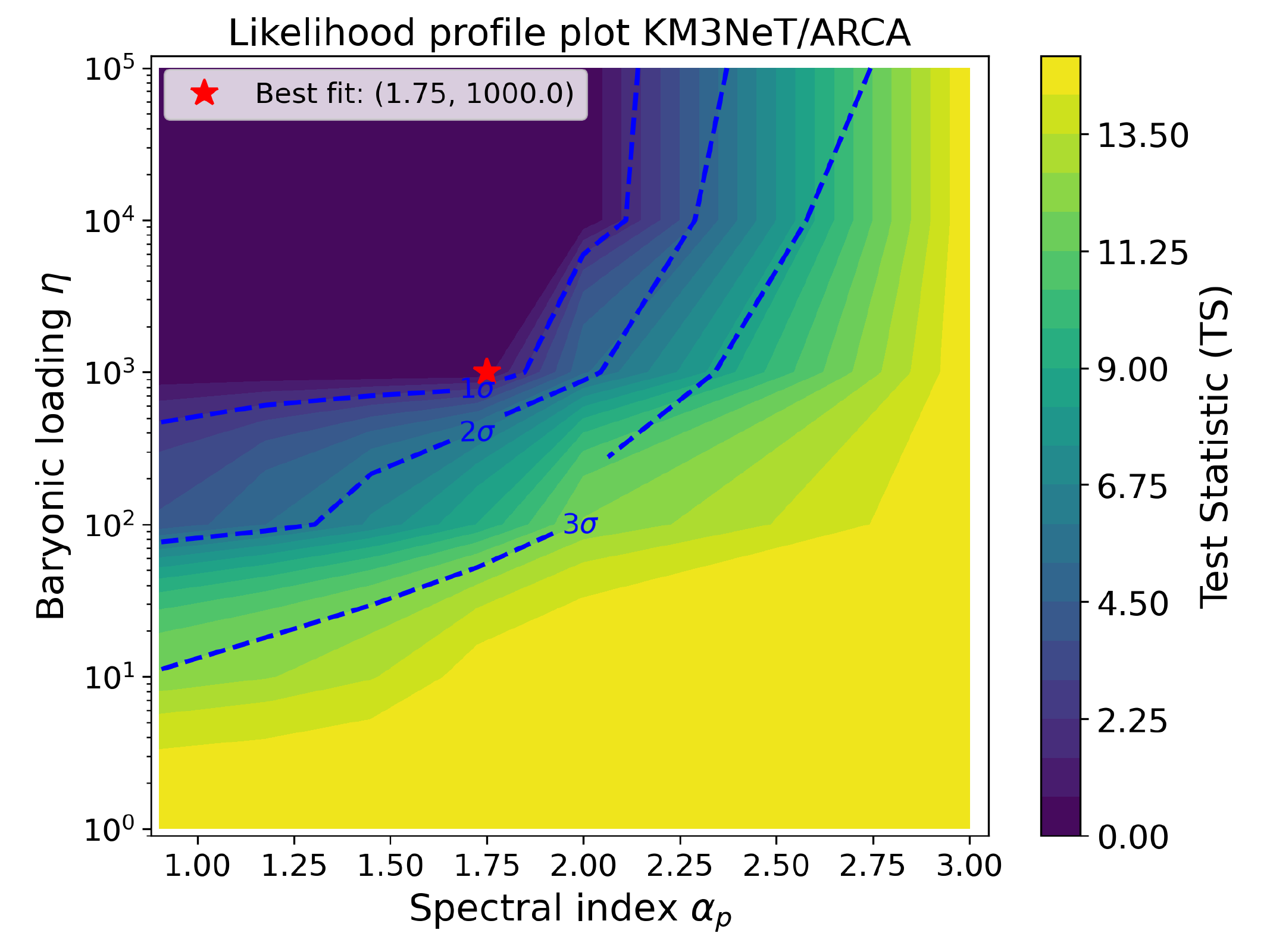}
    \caption{Test statistic contour plot in terms of baryonic loading~($\eta$) and the proton spectral index~($\alpha_p$) for the KM3NeT/ARCA-only analysis. The best-fit value is reported with a red star.}
    \label{contour}
\end{figure}
The $1,2$ and $3$ $\sigma$ TS contours are shown in Fig.~\ref{contour}. In this case, the best-fit value is $\eta \approx 10^3$ and $\alpha_p \approx 1.75$.  The best-fit spectral index is close to the value obtained when the IceCube exposure is included in the likelihood analysis.
However, in this case the baryonic loading value maximizing the likelihood is $\sim 2$ orders of magnitude higher than for the joint analysis. The two results are compatible within about $2.5$–$3$ $\sigma$.

  
\begin{figure}[h!]
    \centering
    ~~~~~~\includegraphics[width=\linewidth]{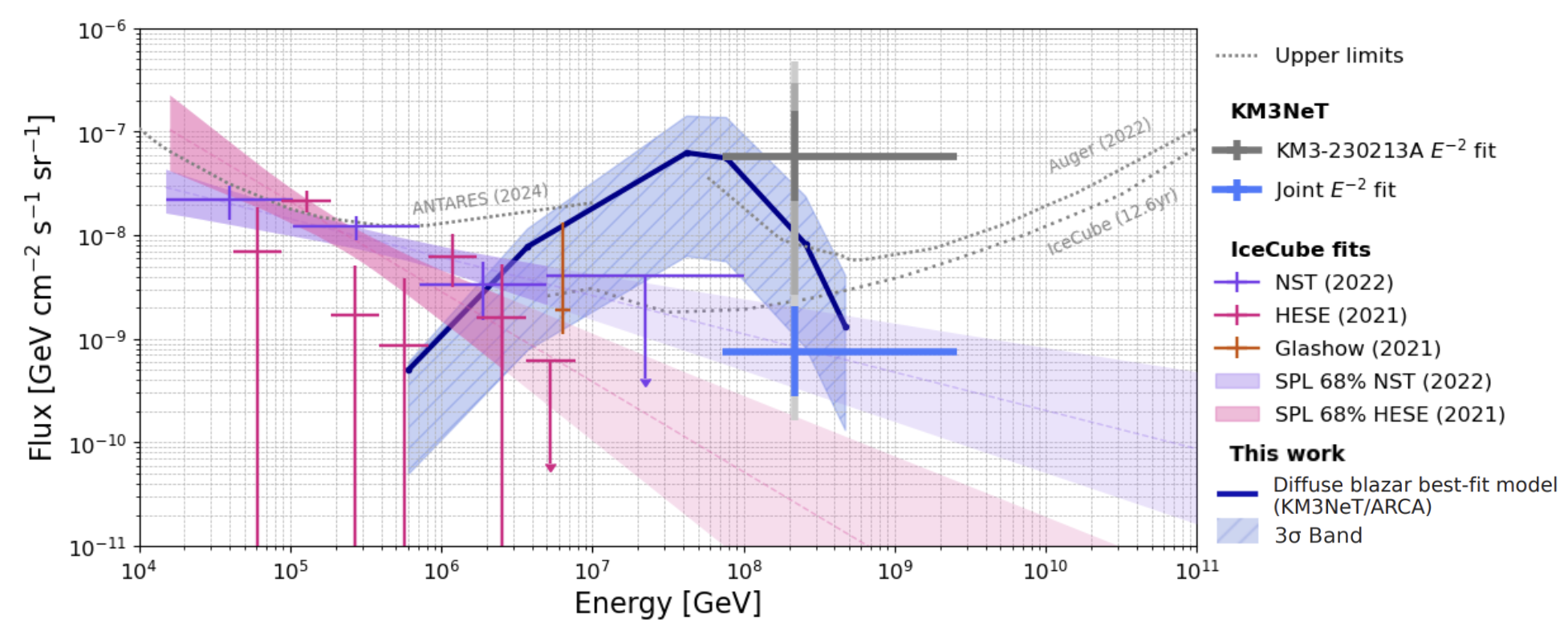}
    \caption{Neutrino  diffuse spectral energy distribution  for blazars in terms of the energy for a single neutrino flavour. The dark blue line represents the best fit while the shaded region is the $3\sigma$ band. The prediction is compared with the KM3-230213A equivalent flux~\cite{KM3NeT:2025npi}, the joint $E^{-2}$ flux obtained by~\cite{KM3NeT:2025ccp} including IceCube-Extreme High-Energy~\cite{IceCube:2018fhm} and Auger non-observations, and the updated IceCube~\cite{IceCube:2025ezc} and Auger differential upper limits~\cite{AbdulHalim:2023SN}. For comparison, the diffuse neutrino flux measured by the IceCube Neutrino Observatory with several samples~\cite{IceCube:2020wum,Abbasi:2021qfz,IceCube:2021rpz}and also the ANTARES upper limits~\cite{ANTARES:2024ihw} are reported. The pink and purple shaded regions represent the IceCube single-power-law (SPL) fits for  High-Energy Starting Events~(HESE)~\cite{IceCube:2020wum} and Northern Sky Tracks~(NST)~\cite{Abbasi:2021qfz}, respectively.}
    \label{Fig:neutrinos_only_KM3}
\end{figure}

\begin{figure}[h!]
    \centering
    \includegraphics[width=\linewidth]{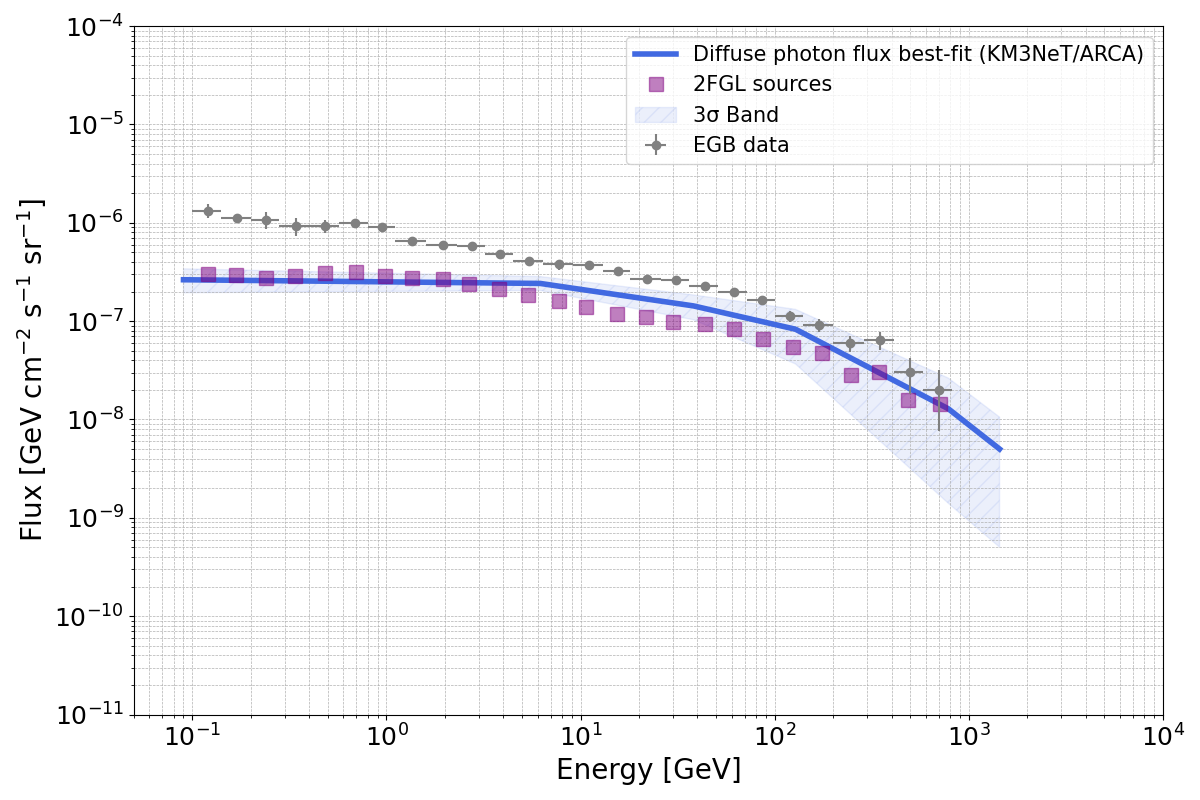}
    \caption{Gamma-ray diffuse spectral energy distribution  for blazars as a function of the energy. The solid line represents the best fit while the shaded region represents the $3\sigma$ band. The result is compared with ExtraGalactic Background (EGB) measurements of Fermi-LAT and the 2FGL sources~\cite{Fermi-LAT:2014ryh}.}
    \label{Fig:gamma_only_KM3}
\end{figure}

The best fit and $1\sigma$ bands for neutrinos and photons are shown in Figs.~\ref{Fig:neutrinos_only_KM3} and \ref{Fig:gamma_only_KM3}, respectively, and are compared with the corresponding measurements discussed in the main text. The neutrino spectrum is compatible with the one reported by~\cite{KM3NeT:2025npi} although it is in tension with the IceCube measurements reported in the $\sim 100-1000\, \rm TeV$ energy range. Furthermore, the flux is in tension with the Auger and IceCube upper limits at energies above $\sim 100\, \rm PeV$. This suggests that more luminous sources are required to enable the detection of $\sim 1$ event in KM3NeT/ARCA. However, the corresponding gamma-ray flux is consistent with Fermi-LAT constraints; in fact, the fraction produced by blazars is $\simeq 77\%$, as shown in Fig.~\ref{Fig:gamma_only_KM3}.




\end{document}